\documentclass[prd,nofootinbib,twocolumn,showpacs]{revtex4}
\usepackage{graphics,rotating,multirow,bm,color,hyperref}
\hypersetup{
    colorlinks=true,
    linkcolor=red,
    citecolor=blue,
}

\def\etal{{\frenchspacing\it et al.}}

\def\eg{{\frenchspacing\it e.g.}}

\def\be{\begin{equation}}
\def\ee{\end{equation}}
\def\ba{\begin{eqnarray}}
\def\ea{\end{eqnarray}}
\begin{document}

\title{Investigating dark energy experiments with principal components}

\author{
Robert G. Crittenden$^1$, Levon Pogosian$^{2,3}$, and Gong-Bo Zhao$^{1,2}$}

\affiliation{$^1$ Institute of Cosmology and Gravitation, University of Portsmouth,
Portsmouth, PO1 2EG, UK \\
$^2$Department of Physics, Simon Fraser University, Burnaby, BC, V5A 1S6, Canada \\
$^3$ Department of Physics, Syracuse University, Syracuse, NY 13244-1130, USA
}

\begin{abstract}
We use a principal component approach to contrast different kinds
of probes of dark energy, and to emphasize how an array of probes can work together to constrain an arbitrary equation of state history $w(z)$. We pay particular attention to the role of the priors in assessing the information content of experiments and propose using an explicit prior on the degree of smoothness of $w(z)$ that is independent of the binning scheme.  We also show how a figure of merit based on the mean squared error probes the number of new modes constrained by a data set, and use it to examine how informative various experiments will be in constraining the evolution of dark energy.
\end{abstract}

\date{\today}

\maketitle

\section{Introduction}

There is growing evidence indicating that the expansion rate of the
Universe is accelerating, either due to modified gravitational
physics or some new type of repulsive `dark energy' coming to
dominate the Universe.   Evidence for this comes from several
directions. First, measures of the present dark matter density show
that it is unable to explain the observed Hubble expansion rate
\cite{2dFSDSS,hconstraint}. Second, probes of the past expansion
rate using high redshift supernovae (SN) as standard candles
directly show acceleration \cite{RiessPerl,Kowalski:2008ez}, and this is confirmed
by the angular size of features in the cosmic microwave background
(CMB) \cite{WMAP} and the baryon acoustic features in the galaxy
power spectrum \cite{BAO}.
Finally, CMB-large scale structure correlations (see \eg~ \cite{Cross}) provide evidence that the growth rate of fluctuations, which depends directly on the background expansion history, is inconsistent with a matter dominated universe.

Studies are underway to improve the observational picture: SN
surveys will be expanded and pushed to higher redshifts, microwave
background and cluster studies will improve the constraints on the
present matter density, and gravitational lensing and high redshift
large scale structure surveys will probe directly the growth rate.
These will seek to answer some fundamental questions: Are the data
consistent with a cosmological constant, or is there evidence for
some kind of dynamical dark energy, such as quintessence?  Are the
direct probes of the background expansion history consistent with
the growth of perturbations for some dark energy model, or is a
modification of gravity required?

These questions would be straight forward to answer given a specific
dynamical dark energy model, or even a family of such models.
However, there is no favored dynamical model.  Even in the
quintessence models, where the dark energy is due to a scalar field
rolling down a potential well, virtually any behavior of the
equation of state $w(z)$ is possible by choosing the appropriate
potential \cite{CaldStein}. This makes it essentially impossible to
find an experiment which will conclusively prove that the dark
energy is or is not dynamical.  Even if we find that data are
consistent with a cosmological constant, it is very likely that
there would still exist a family of dynamical DE models which would
improve the fit to the data.  The question would remain whether any
of these models is expected on theoretical grounds.

Another important question is whether the constraints from measures of $w(z)$ coming from the background expansion (CMB, SN) are consistent with those arising from the growth of perturbations (correlations, weak lensing), since inconsistencies could potentially indicate a breakdown of general relativity on large scales \cite{IUS05}.  If observations appear inconsistent, it might be explained by a high frequency feature in $w(z)$; again, we must rely on theory to determine how baroque this solution is, or whether modified gravity is more natural.

Given this fundamental difficulty, we choose to focus instead on the observations and
what they might potentially tell us, using a principle component approach introduced in the
dark energy literature by Huterer and Starkman (HS) \cite{HutStark} and subsequently used by \cite{KAS}.
Any study of dynamical dark energy models unfortunately must begin with some kind of
 parameterization which implicitly
imposes a measure on the space of models.  For this discussion we focus on
parameterizing $w(z)$ alone, though in principle we might also include the dark energy
sound speed.

Choosing a parameterization is necessarily arbitrary and answers depend on the
choice \cite{Bruce04,LH05}.  We choose our basis following a few simple principles.
First, it should have enough freedom to be able to reproduce most of the
phenomenological $w(z)$ used in the literature, as well as the $w(z)$ typically derived
by the potentials which have been considered
(e.g. \cite{WellerAlbrecht}). Second, we would like to keep as few parameters as possible, as
long as they are capable of capturing the effect of $w(z)$ on the observations.
To implement these, we first assume that $w(z)$ is reasonably smooth and bin in
finite redshift bins which implicitly excludes variations on time-scales smaller than
a bin width.
We also cut off the coverage at high redshift, since no experiment is likely to constrain $w(z)$
at these redshifts (unless $w(z)$ becomes larger than zero at some point.)

We choose 40 uniform redshift bins, stretching to a maximum redshift of $z=3$.
We have checked that our results are relatively independent of
the size of the bin and whether we use a linear or logarithmic binning.
The results are also little affected if we allow the high redshift ($z>3$) equation of state to vary or if we fix it at the fiducial value.
To avoid $w(z)$ with infinite derivatives, each bin rises and falls following a hyperbolic tangent
function with typical $dz$ of order 10\% of the width of a bin.  As the fiducial model, we choose a constant $w_0 = -1.0$ model;
this is not only consistent with the present data, but also can be
used as a fiducial model for most other phenomenological
parameterizations considered so far. Since dark energy
perturbations (DEP) play a crucial role in the parameter
estimation~\cite{WellerLewis, ZhaoWMAP3}, we use a modified version of
CAMB which allows us to calculate DEP for an arbitrary $w(z)$
consistently~\cite{DEP}.

For the principle component analysis, we calculate the Fisher matrix
based on four kinds of experiments: supernovae surveys, CMB
anisotropies, galaxy counts (GC) correlation functions and weak
lensing (WL) observations; we also include possible
cross-correlations between these, including CMB-galaxy and CMB-weak
lensing (which are sensitive to the integrated Sachs-Wolfe effect)
as well as galaxy-weak lensing correlations.   We consider the
limits which might be obtained in a decade's time, by experiments
like JDEM~\cite{JDEM}, Planck~\cite{Planck} and LSST~\cite{LSST}.

We follow the HS conventions for the principle component analysis, but our underlying approach is Bayesian rather than frequentist.   We first calculate the Fisher matrices for each of the experiments, $F^{a}_{ij}$, where the indices $i,j$ run over the parameters of the theory, in our case the binned $w_i(z)$.  We find the normalized eigenvectors and eigenvalues of this matrix ($e_i(z),\lambda_i$), and so can write
\begin{equation}
F = W^T \Lambda W,
\end{equation}
where the rows of $W$ are the eigenvectors and $\Lambda$ is a diagonal matrix with elements $\lambda_i$.
The Fisher matrix is an estimate of the inverse covariance matrix we expect the data to give us and the
eigenvalues reflect how well the amplitude of each eigenvector can be measured.  The true behavior of the equation of state can be expanded in the eigenvectors as
\begin{equation}
w(z) = w_{\rm fid}(z) + \sum_{i=1}^N \alpha_i e_i (z)
\end{equation}
and the expected error in the recovered amplitudes is given by $\sigma(\alpha_i) = \lambda_i^{-1/2}$ (assuming the true model is reasonably close to the original fiducial model which was used to calculate the Fisher matrix.)

\section{Choosing a prior}

Lacking any prior knowledge of the possible $w(z)$ functions, all of the eigenvectors are informative, no matter how large the error bars.  However, this is unduly pessimistic.  Even without a physical model for $w(z)$, we would still be surprised if it was much too positive ($w \gg 1/3$) or much too negative ($w \ll -1$).  Such preconceptions constitute our theoretical priors, and in principle allow us to roughly separate the eigen modes into those which are informative relative to the priors and those that are not.  If the error bars on an eigenmode allow much too positive or too negative $w(z)$, then we have not really learned anything.

We need some way of quantifying our theoretical biases; however, this is necessarily
subjective and begs theoretical input.  One possible choice is to use a weak Gaussian prior on the amplitude of $w(z)$ in any given bin, e.g. $w_i = w_0 \pm \sigma_p,$ and assume that the bins are uncorrelated.
This would prevent $w(z)$
from deviating too much from the fiducial model in any given bin.
The problem is that in this case formally all the modes have the same
importance which does not reflect our actual prior assumptions.
The choice to
bin the parameter is implicitly motivated by an assumption that the parameter is in some
sense smooth, that high frequency modes are less likely or less interesting than low
frequency modes.
A binning in effect provides a sharp cutoff, giving equal weight to all modes below the Nyquist
frequency and no representation of modes with higher frequencies.  It is perhaps more intuitive
that the prior should have a more gentle transition, whereby the prior probability of a mode
is gradually decreased as its frequency increases.

One alternative way of implementing priors on $w(z)$ is to choose a correlation function describing fluctuations away from some fiducial model.
Lacking a specific prior from fundamental theory, we propose treating the equation of state
as a random field evolving with a given correlation period (in time or redshift.)   Much like the choice to bin in the first place, this prior is based
on the assumption that the equation of state is evolving smoothly.   While in this paper we focus on using correlations in redshift, a more physical independent variable could be used, such as scale factor or proper time.   Since the prior is really only an extension of the binning choice, ideally one should use the same independent variable for both.

In effect, the correlation function prior provides a transition
between high frequency oscillations, which are resisted by the
prior, and the low frequency modes, which are unaffected. Providing
a prior stabilizes the high frequency variances and allows us to
focus on the more interesting low frequency modes. Also, as long as
there are sufficient bins compared to the correlation length, the
prior largely wipes out dependence on the precise choice of binning.

In practice, we need to construct the covariance matrix associated with the prior (which will be inverted
and added to the Fisher matrix from the observations.)
The starting point is to assume that the deviations of the equation of state from its
fiducial model (e.g. $w= -1$) can be encapsulated in a correlation function:
\begin{equation}
 \xi_w (|z - z'|) \equiv \left\langle (w(z) - w_{\rm fid}(z))(w(z') - w_{\rm fid}(z')) \right\rangle.
\end{equation}
Such a form implicitly assumes independence of translations in redshift; that is, that there is
no preferred epoch for variations from the fiducial equation of state.
(Though such a preferred epoch could be built into the fiducial model itself.)   As stated above, we choose
redshift as our independent variable, but the same considerations would apply to another choice,
such as the scale factor.

Let us assume the i$^{th}$ bin is from $z_i$ to $z_{i} + \Delta$, and for simplicity we
will assume that all bins have the same width $\Delta = z_{i+1} - z_i$.
The equation of state averaged over each bin is given by
\begin{equation}
w_i = \frac{1}{\Delta} \int_{z_i}^{z_i+\Delta} dz \, w(z).
\end{equation}
We can write the variation from the fiducial model averaged over the bin as, $\delta w_i =  w_i - w_{fid}$.
Calculating the covariance matrix of the binned equation of state is then straightforward:
\begin{eqnarray}
\langle \delta w_i \delta w_j  \rangle =  \frac{1}{\Delta^2} \int_{z_i}^{z_i+\Delta} dz  \int_{z_j}^{z_j+\Delta}  dz' \, \xi_w (|z - z'|) .
\end{eqnarray}
All that remains is to calculate this covariance matrix for a given functional form of the correlation function.

We assume that it has a characteristic correlation redshift distance $z_c$ after which it falls off;
for example, let us assume $\xi_w (z) =  \xi_w (0) /(1 + (z/z_c)^2).$  In this case we can perform the integrals analytically,
using the relations,
\begin{equation}
\int \frac{dz}{1 + z^2/z_c^2} = z_c \tan^{-1} {z \over z_c}
\end{equation}
\begin{equation}
\int dz \tan^{-1}{z \over z_c} = z \tan^{-1}{ z \over z_c} - \frac{z_c}{2} \log \left(1 + {z^2 \over z_c^2}\right).
\end{equation}
The covariance between two bins of width $\Delta$ separated by $\tilde{z} = |z_i -z_j|$ can be shown to be
\begin{eqnarray}
\langle \delta w_i \delta w_j \rangle & =&   \xi_w(0) \frac{z_c^2}{\Delta^2} [  x_{+} \tan^{-1}x_{+} +  x_{-} \tan^{-1}x_{-}
 \nonumber \\ & & - 2 \bar{x} \tan^{-1} \bar{x}  + \log (1 + \bar{x}^2)
 \nonumber \\ & & - \frac{1}{2} \log (1 + x_{+}^2)  - \frac{1}{2} \log (1 + x_{-}^2)  ]
 \label{eq:covariance_ij}
\end{eqnarray}
where $\bar{x} \equiv \tilde{z}/z_c$, $x_{+} \equiv (\tilde{z}+ \Delta)/z_c$ and $x_{-} \equiv (\tilde{z} - \Delta)/z_c$.
Once the bin width and the correlation length are set, the correlation matrix will only depend on $\tilde{z},$ the distance between the bins of interest.

Note that the variance of the mean equation of state over all the
bins follows directly from this by taking $\Delta$ to be the entire
redshift interval and $\tilde{z} = 0$.  As long as $z_c/z_{\rm max}
\ll 1$, one can show that
\begin{equation}
\sigma_m^2 =  \int_0^{z_{\rm max}} {dz} \int_0^{z_{\rm max}} {dz'
\over z_{\rm max}^2}\xi_w(z-z') \simeq {{\pi} \xi_w(0) z_c \over z_{\rm
max}}.
\end{equation}
This is essentially the variance at any given point, $\xi_w(0)$,
divided by the number of effective degrees of freedom, $N_{\rm eff}
= z_{\rm max}/z_c$.   While this expression is appropriate in the
$z_c/z_{\rm max} \ll 1$ limit, in practice we use the full expression,
since the corrections can be significant for larger correlation
lengths ($z_c \ge 0.4$).

In practice, our inputs to the prior were the error in the mean, $\sigma_m$, and the correlation distance, $z_c$.   We chose the error in the mean to be of order $\sigma_m \sim 0.2-0.5$, which seemed representative of our present observational uncertainty in the mean.  We tried correlation lengths in the range $0.1 \le z_c \le 0.4$, where the upper limit was beginning to be strong enough to impact the observed modes for the SN.  The SN results were the first affected because they are the only probe we considered which constrained (however weakly) the shorter wavelength DE modes.

The resulting prior takes the form,
\begin{equation}
{\cal{P}}_{\rm prior} \propto \exp{\left(-\frac{1}{2}({w}_i^{\rm
true} - {w}_i^{\rm fid})C_{ij}^{-1}({w}_j^{\rm true} - {w}_j^{\rm
fid})\right)}
\end{equation}
where $C_{ij} \equiv \langle \delta w_i \delta w_j \rangle$ is the correlation function $\xi_w (|z - z'|)$ integrated over the bins (Eq.~(\ref{eq:covariance_ij})).  This prior naturally constrains the high frequency modes without over constraining the lower frequency modes that are typically probed by experiments.

As an aside, note that assuming no bin correlations, i.e., a purely
diagonal matrix for the prior, is equivalent to using a delta
function for the correlation prior, e.g. $\xi(z) = \xi_0 \delta(z).$
(Any finite correlation distance will automatically generate some
off-diagonal correlation.)   In such a case, one finds $\langle
\delta w_i \delta w_j \rangle = \xi_0 \delta_{ij}/\Delta $  and the
mean variance is $\sigma_m^2 = \xi_0 /z_{\rm max}$.   Thus, assuming
a fixed total range, the bin variance should grow with the number of
bins $\langle \delta w_i^2 \rangle = \sigma_m^2 N_{\rm bins}$ to
keep the mean variance unchanged.

\section{Comparing dark energy probes using PCA with the smoothness prior}

\subsection{Figures of merit}

In order to compare the information content of various probes, one
needs to decide on a so-called Figure of Merit (FOM): a scalar
quantity that one should optimise. The most often used scalar
quantity relates to the determinant of the Fisher or curvature
matrix, which is effectively a measure of the volume of the
parameter space. For example, this (or its square root) is used for
the DETF figure of merit. There are good reasons for trying to
minimize the volume of phase space, particularly in the context of
comparing the Bayesian evidences of different models.  In these
calculations, an Occam's factor plays a role, which is basically the
ratio of the total possible volume of parameter space to the volume
allowed by the data.   Thus minimizing this volume factor would
allow us to rule out models with higher significance.

While the parameter space volume is a very natural measure, it is
not the only possible measure. Another possibility is to focus on
the trace of the inverse Fisher matrix, which corresponds to
minimizing the sum of the variances, also known as the mean squared
error (MSE). This is dominated by the least constrained modes, so
minimizing it will tend to spread the information out among many
different eigenvectors.  This is useful for a number of reasons. One
is to minimize the errors in reconstructing the true behavior of the
parameter from observations, which is directly quantified by the
MSE.   More modes also provide the opportunity for checks of the
consistency of the data.   If the number of well constrained modes
is less than the number of parameters in our model, there will
always exist degeneracies in the model; this makes it impossible to
check the consistency of data within that model.

For example, in the one-parameter  $\Lambda$CDM model, a single dark energy observation is sufficient to
constrain the DE density and we could use the CMB acoustic scale to constrain it. If another observation, such as a single well-measured SN eigenmode, is available, we can check that it gives a consistent answer for that one parameter.  But for a two parameter theory (e.g. constant $w$ or $\Lambda$ with curvature)  these data must be combined to find a unique model, and no longer can be used as a consistency check.  And for a three
parameter theory (like linear evolving $w(z)$), more SN eigenmodes or other types of data must be found to find a unique set of model parameters. Thus, it pays to constrain more modes than your theory has parameters.


It is straightforward to use the MSE as a FOM to evaluate the value added by a given experiment.
Ideally, we want to minimize the difference between the true and estimated $w(z)$,
\begin{equation}
{\rm MSE} \equiv \sum_i (w^{\rm est}_i- w^{\rm true}_i)^2 = \sum_i
(\alpha^{\rm est}_i- \alpha^{\rm true}_i)^2,
\end{equation}
the latter following from the orthonormality of the eigenvector basis.  In the absence of any priors, this is
expected to be
\begin{equation}
\langle {\rm MSE} \rangle = \sum_i \sigma(\alpha_i)^2 = {\rm Tr}
\,F^{-1}.
\end{equation}
This will be dominated by the poorest determined modes, which is why
having some kind of prior is essential. Taking into account priors,
$F$ is replaced in this expression by $C^{-1}+F$. The MSE arising
from the prior alone is ${\rm MSE} = {\rm Tr}\,  C = N
\bar{\xi}_w(0) $ where $\bar{\xi}_w(0)$ is the variance in a single
bin; this is independent of the shape of the correlation function.
Adding more experimental data reduces the MSE
as the well determined modes contribute less. The amount by which
the MSE is reduced can be seen as a measure of how informative the experiment is.

For the MSE criteria, it will generally be more effective to maximize the number of modes which are reasonably well determined ($\sigma(\alpha_i) < \sigma_p$) than to have a smaller number of better determined eigen modes.  To reduce the MSE, it is most effective to focus on the modes which have larger error bars.
Much different conclusions can result from using a different FOM \cite{Bassett_etal}.
For example, if we used the volume of the parameter error ellipses, rather than the
MSE, as the figure of merit, a factor of two reduction in any error bar would lead to the same reduction in the volume, regardless of how well determined the error bar was initially.   If the parameter was already tightly constrained, the volume could be reduced substantially while the MSE would be largely unchanged.

\subsection{Application to future data}


For our forecasts, we assume the following probes: LSST for WL and
GC, Planck for CMB, and a Joint Dark Energy Mission (JDEM) for SN.
For the galaxy distribution we assume the most optimistic LSST-like
survey with several billion galaxies distributed out to $z=3$. We
then divide the total galaxies into ten and six photometric bins for
the calculation of GC and WL respectively. The survey parameters
were adopted from the recent review of the LSST
collaboration~\cite{LSST2}. Namely, we use $f_{\rm sky}= 0.5$, $N_G
= 50$ gal/arcmin$^2$ for both WL and counts; the shear uncertainty is assumed to be $\gamma_{\rm rms} =
0.18 + 0.042 z$, and the photometric redshift uncertainty is given by $\sigma(z) = 0.03 (1 + z)$. We only use the
information from scales that are safely in the linear regime
(corresponding to $k\le 0.1$h/Mpc). For CMB we include the
Planck temperature and polarization spectra and their cross-correlation; for the SN we 
assume the detection of 2000 SN distributed out to a redshift of 1.7.  
Details of the assumptions for the experiments and the calculation
of the Fisher matrices can be found in \cite{PSCCN} and in
\cite{WL}.

The eigenvectors and eigenvalues clearly depend on how we treat the
other cosmological parameters.  For the CMB constraints, we use the
CMB data alone, assuming a flat universe, and marginalize over other
cosmological parameters including the dark matter density, baryon
density, spectral index, Hubble parameter, and optical depth.  For
the SN, we marginalize over the dark matter density and the value of
the intrinsic SN magnitude $M$, but since the dark matter density is
likely to be well determined, we use the CMB Fisher matrix
marginalized over dark energy parameters as the prior. Formally, one
should add instead the full CMB and SN Fisher matrices to find out
the joint dark energy eigenmodes; however, by first marginalizing
over the CMB dark energy parameters, one obtains a clearer picture
of what the SN are actually measuring, as otherwise the first CMB
mode would contaminate the SN information. We do similarly for the
GC and WL auto-correlations, but for GC, we marginalize
also over a different possible bias in each of the ten photometric
redshift bins.  (The assumption of independent biases is conservative and significantly reduces the
effectiveness of the GC spectra to tell us about dark energy.)
Finally, for the cross-correlations, we use the CMB, WL and the
GC data to give priors for the bias and other cosmological
parameters.

\begin{figure}[thp]
\centering
{\includegraphics[scale=1.13, ]{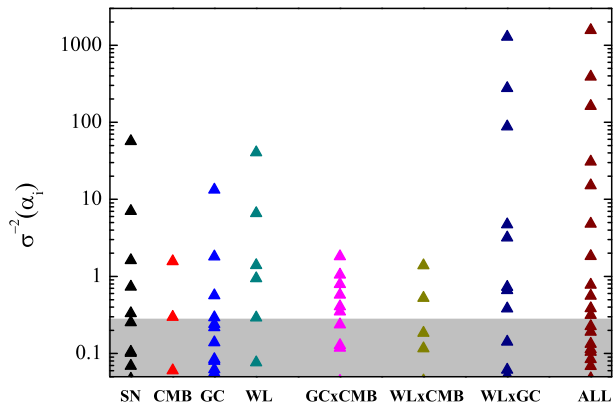}}
\caption{The eigenvalues ($\sigma^{-2}(\alpha_i)$) for the raw Fisher matrices (no priors assumed) for
different experiments. The grey shaded region shows the diagonal
prior of $\sigma_m\geq0.3$.
}
\label{fig:eval}
\end{figure}

Fig.~\ref{fig:eval} shows the spectra of eigenvalues for the
various data sets, their cross correlations and the combined data.
The shaded region provides a rough threshold based on a diagonal
prior with $\sigma_m\geq0.3$, and any eigenmodes above this are
taken to be informative.  Many of the experiments provide multiple
independent modes, with upwards of ten informative modes for the
combined data set. For a correlated prior, the threshold to be
informative depends on the frequency of the mode and this can mean
that fewer informative modes are found.

\begin{figure}[thp]
\centering
{\includegraphics[scale=0.62, ]{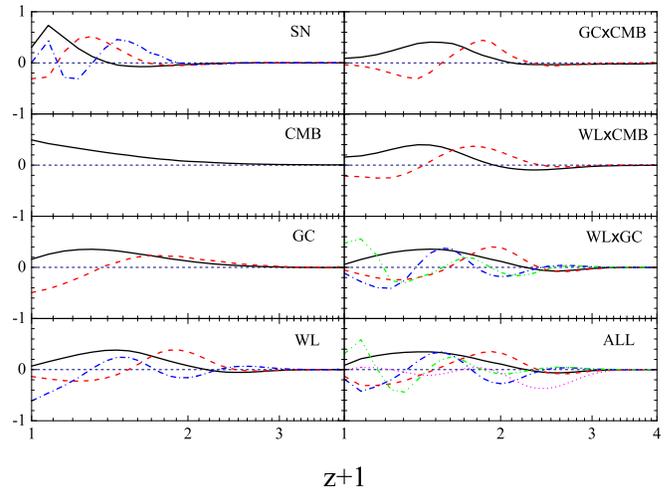}}
\caption{The best determined eigenvectors for different kinds of experiments. No priors have been assumed and the amplitudes are normalized to unity. Only the first five modes are plotted if more than five modes are well-constrained. The modes are
shown, in the order from better constrained to worse, as black solid, red dashed, blue dash-dot, green dash-dot-dot and magenta dot curves. With additional modes, we can begin to probe higher frequency changes in $w(z)$; however, all the data is primarily constraining only the low redshift behaviour ($z<1$).
}
\label{fig:evec}
\end{figure}

The best determined eigenvectors are shown in Fig.~\ref{fig:evec}
for the various types of measurements we consider.  We plot only
those with eigenvalues above a given threshold.  The predictions
vary significantly depending on the experiment.  The CMB gives one
well determined mode, which corresponds to the angular distance to
the last scattering surface \cite{sarah03}.  The SN and the GC
correlations give a larger number of well determined modes,
which is a benefit of the range in redshifts of the data.  The
cross-correlation only gives a single marginally determined mode,
but it probes to higher redshifts than many of the others. Finally,
we show the well determined eigenvectors for the total Fisher
matrix.  When combined, the experiments can probe higher frequency
modes than they can probe independently. This is because the
combination is sensitive to the differences in the individual
eigenvectors.
As might be expected from our choice of fiducial model ($w_0 = -1$), the experiments are most sensitive to lower redshifts.

How much the experiments improve the MSE depends on the choice of correlation function for the prior.
In Table 1, we vary $\xi_w(0)$ and $z_c$ while holding the prior constraint on the mean fixed, $\sigma_m = 0.3$.
If we choose small $z_c$, all the bins are effectively uncorrelated and the variance in each bin is large.   In this case, the MSE is large and all of the eigenmodes are constrained by the prior at the same level.  As we let $z_c$ get larger, the variance in each bin shrinks, as does the MSE.  Modes with wavelengths smaller than $z_c$ become tightly constrained by the prior.

For the case of a diagonal prior, these numbers are easy to
interpret in terms of the number of newly constrained modes.  For
the prior alone, the MSE is given by $N_{\rm bins} \times \langle
\delta w^2 \rangle = N_{\rm bins}^2 \sigma^2_m = 144.$   Adding the
forecast data, some of the modes will have significantly reduced
errors bars compared to the prior alone.  As the prior is diagonal,
the new eigenmodes were also eigenmodes of the prior with the same
variance, given by  $N_{\rm bins} \sigma^2_m = 3.6.$   The reduction
of the MSE thus roughly tells us how many new modes are more
constrained compared to the prior.  The SN data, for example, reduce
the MSE from 144 to 125, which means there is new information  on
$(144-125)/3.6 \simeq 5 $ modes, beyond what was assumed in the
prior.  This can be seen in the spectrum of eigenmodes, shown in
Fig.~\ref{fig:eval},  counting the number of modes with amplitudes
above the prior threshold shown by the horizontal line.

For a correlated prior, the interpretation of the MSE is less
straightforward for two reasons.  First, the prior eigenvalues are
no longer all the same; as some start out higher, improving
information on these modes has less of an impact on the total MSE
than in the diagonal case.  More fundamentally, in this case the eigen vectors change
when the forecast data are included, so it is no longer
possible to have a one-to-one correspondence of modes with and
without the forecast data. We can, however, put a lower bound on the
number of new modes by dividing by the largest prior variance; this
will generally be for the homogenous mode, which as above is set by
the constraint on the variance of the mean  ($N_{\rm bins}
\sigma^2_m$). As can be inferred from the table, the number of new
modes estimated in this way is significantly reduced, especially as
the prior correlation length is increased.

The relative value of different experiments depends somewhat on what was assumed to be known already.
For small $z_c$, the SN experiments are one of the most informative of the single data set we consider.  However, as $z_c$ is increased, the higher frequency modes probed by the SN are already strongly constrained by the prior, reducing the impact of the SN data; other observations, such as weal lensing, measure lower frequency modes, and are not as affected by the change in the prior.

\begin{table}
\begin{center}
\begin{tabular}{@{}l@{~~~}c@{~}c@{~}c@{~}c@{}c@{}}

                        & diagonal  & $\,\,z_c=0.1\,\,$  & $\,\,z_c = 0.4$ \,\,\\
\hline
no data                   & 144.0 & 35.0  & 11.5 \\
\hline
SN                        & 124.9 \, [5.3] & 27.4  & 8.2 \\
CMB                       & 138.4 \, [1.6] & 31.0  & 8.6 \\
GC                        & 126.1 \, [5.0] & 26.3  & 6.7 \\
WL                        & 128.5 \, [4.3] & 24.7  & 5.6 \\
GC$\times$CMB             & 124.9 \, [5.3] & 24.7  & 7.2 \\
WL$\times$CMB             & 136.1 \, [2.2] & 29.9  & 8.9 \\
WL$\times$GC              & 117.3 \, [7.4] & 21.3  & 5.1 \\
\hline
Total                     & 102.6 \, [11.5] & 16.7  & 3.0\\
\hline
\end{tabular}
\end{center}
\caption{The mean squared error (which is related to the number of well constrained modes) for various priors
and experiments. The priors are normalized so that $\sigma_m =0.3$
and the `diagonal' prior means no correlations exist between bins.  For this diagonal prior, we give in brackets the inferred number of
modes meaningfully constrained by the observations.
\label{tab:mse}}
\end{table}

\section{Additional applications of the PCA method}

\subsection{Reconstructing w(z)}

Given the measurements, one would like to reconstruct a best estimate of the equation of state history $w(z)$.   In our Bayesian approach,
this simply means finding the model parameters which maximize the posterior probability distribution, which is the prior distribution times the likelihood expected from the observational Fisher matrix.  Some of the eigen modes will be determined by the data, and some will be determined by the prior.   In particular, if the Fisher matrix error for an eigenmode, $ \sigma(\alpha_i)$, is much smaller than that from the prior, the parameter will be well determined and independent of the prior assumptions.  Modes where $\sigma(\alpha_i)$ from the observational Fisher matrix are much larger than the prior error will be poorly determined and the amplitudes will revert to their value at the peak of the prior, which is the fiducial model.

Earlier discussion of PCA's took a more frequentist approach to reconstructing $w(z)$ \cite{HutStark}, where the function was
reconstructed by using a subset of the PCA modes.  Deciding how many modes are kept requires minimizing a `risk function' which is
effectively the mean squared error, separated into a variance and a bias contribution.   The size of the bias depends on how much the
true underlying model differs from the fiducial model which is assumed; thus the number of modes one keeps depends on
what you think the underlying model is, which is obviously unknown.    Thus, there is some ambiguity in the prescription; it would however
make sense to base the bias not on a single model, but on the ensemble of potential models.   This is precisely what we attempt to
quantify with our choice of Bayesian prior.

Not too surprisingly, the Bayesian and frequentist approaches should yield similar reconstructions.   The modes excluded from the
frequentist estimator are precisely those modes where the Bayesian prior overwhelms the information from the observations.  In regions where
the data are good, the true model will be reconstructed well; in regions where the data are poor, the reconstruction reverts to the fiducial
model assumed on theoretical grounds.

The main difference is that the frequentist estimator explicitly drops the more poorly determined modes from the beginning, which means no
matter how large they are measured to be, they will not affect the reconstruction.   Since these modes are dropped, the reconstructed error
bars for modes where there is no information actually appear smaller, which is clearly incorrect.  In the Bayesian case, however, the errors instead revert to the theoretical uncertainty, which is more representative of our degree of ignorance.

The predicted MSE gives a good idea of the expected errors in the reconstruction, assuming the true model is typical of those allowed by the prior. �In that sense, the MSE is a useful figure of merit. �From the MSE values in Table~\ref{tab:mse}, one can see that the reconstruction will be very poor in the case of the diagonal prior we have assumed. �This is because the prior knowledge we assumed was weak, and even after the addition of data, much uncertainty remains (i.~e. MSE is still large). �The correlated (non-diagonal) prior gives a much smaller MSE, and so should give a much better reconstruction assuming the true model is typical given the prior (e.~g. does not have high frequency oscillations.) �This seems to be a significant advantage of the correlated prior. �

\subsection{PCA for data compression}

A key advantage to the PCA approach is that it retains information about all the modes which the observations are able to constrain, and
thus can be used as a compressed form of data which can be applied to any theoretical parameterization that can be translated into
our binning choice.  Rather than separately calculating the constraints for each parameterization from the original data, this can be done
directly from the principal component representation, and this will apply both to forecasts as well as to the measured constraints.

We can easily emulate any other $w(z)$ parameterization to find the predicted parameter error bars without regenerating the Fisher matrices from scratch.
All we need is to expand the derivative with respect to the given parameter in terms of our eigenmode basis.
That is, for each of the new parameters, we find the coefficients $\alpha^a_i$ such that
\begin{equation}
{\partial {w(z)} \over \partial p^a} = \sum_i \alpha^{a}_i {e}_i(z).
\end{equation}
We can then find the Fisher matrix in the new basis by simple rotation and holding fixed the remaining dark energy parameters
\begin{equation}
F_{ab} = \alpha^a_i F_{ij} \alpha^b_j = \sum_i \alpha^a_i \alpha^b_i \lambda_i
\end{equation}
Obviously it helps if the fiducial model is the same for the various parameterizations, and constant $w$ is a reasonable choice.  
We have checked that this prescription works with a few percent accuracy for the constant $w_0$ parameterization, and for the parameterization linear in the scale factor, and expect the same for any function which can be well approximated by our binning.
The eigen basis is also a useful basis for searching for the peak of the likelihood surface, since the
various amplitudes become decorrelated.

\section{Conclusions}

The choice of prior is a critical one in the study of dynamical dark energy; it tells us how informative a given experiment will be and must also play a role in answering the fundamental questions about whether dark energy is dynamical or whether a modified gravity model should be preferred.   The principal component approach attempts to remove the reliance on particular simplified parameterizations, such as a constant or linearly varying $w(z)$, and replace them with more generic assumptions about the behavior of dark energy.
Here we have focused on a phenomenological approach to the prior, but in principle one wants to base this on theoretical models.  For example, if one had a measure on the possible quintessence potentials, one could translate this into a measure for $w(z)$.

Once a prior is agreed, the principle components and MSE provide a basis for planning a coherent strategy to study dark energy.   The principle components demonstrate any experiment's individual sensitivity and potential for adding information orthogonal to other experiments.  We can then use this information to investigate how effective different experiments or combinations of experiments will be in reducing MSE from its value based on the present data.     The MSE figure of merit naturally focusses on those degrees of freedom we know the least about, resulting in more constrained modes which can provide consistency checks for a theoretical model.

Finally, the principal components represent an effectively lossless means of compressing the observed data, which can then be used to constrain any theoretically motivated dark energy history without repeating the observational analysis.  This is particularly relevant when experiments are sensitive to modes which are orthogonal to the simplest dark energy parameterizations; in such cases, evaluating only the naive dark energy parameters can greatly undervalue what is actually learned in an experiment.


{\bf Acknowledgments:} We thank Bruce Bassett, Bob Nichol, Paul Steinhardt and Jochen Weller
for useful comments and discussions.
LP was supported in part by the NSF grant NSF-PHY0354990,
by funds from Syracuse University and by Research Corporation. LP and GBZ are supported by NSERC and funds from SFU.

{\bf Note:}  This is an extended version of a paper which first appeared in 2005; here we have expanded the discussion to clarify the key aspects of the paper, especially the correlation function prior and using the MSE to quantify the new information provided by experiments.  A number of papers addressing the principal component approach to dark energy have appeared recently \cite{PC-DE} and raised the profile of this approach; in this context, we felt it worth expanding and clarifying our original discussion of these issues.

\end{document}